\documentstyle[12pt,aps,floats,epsf,epsfig]{revtex}

\title{\bf On radiative decay of $Z^{\prime}$ boson}
\author{ G.A. Kozlov
\\
\em Bogolyubov Laboratory of Theoretical Physics\\
\em Joint Institute for Nuclear Research\\
\em Joliot-Curie st., 6, 141980 Dubna, Russia\\
\em e-mail: kozlov@jinr.ru\\
 }
\begin{document}

\maketitle
\begin{abstract}

In the framework of the extended $SU(2)_{h}\times SU(2)_{l}$ model
we study the radiative decay of an  extra gauge boson $Z^{\prime}$,
 $Z^{\prime}\rightarrow\gamma Z$. The data analysis results of the CDF
collaboration at the Fermilab Tevatron allow one to establish not only the lower bound
of the decay probability $Z^{\prime}\rightarrow\gamma Z$ normalized to Drell-Yan
process $Z^{\prime}\rightarrow \mu^{+}\mu^{-}$,
$R(Z^{\prime}\rightarrow\gamma Z/\mu^{+}\mu^{-})>
1.4\times 10^{-5}\cot^{2}\phi$, but also to estimate the mixing angle $\xi$ between
the states $Z^{\prime}$ and $Z$, $\vert\xi\vert < 5\times 10^{-2}$, taking into
account the additional mixing angle $\phi$ related to the extended gauge group
$SU(2)_{h}\times SU(2)_{l}$. The comparison with the decays
 $Z^{\prime}\rightarrow\tau^{+}\tau^{-}$ and
$Z^{\prime}\rightarrow \bar b b$, $Z^{\prime}\rightarrow \bar t t$ based on the
LEP data is also presented.

%\vspace {1 cm}

%PACS 12.38.Aw, 11.15.Kc, 12.38Aw, 12.38Lg, 12.39Mk, 12.39Pn }
\end{abstract}
%\newpage
\vspace {0.5 cm}
\centerline {\bf 1. Introduction}
%\setcounter{equation}{0}

%{\bf 1}.

It is well known that various extensions of the Standard Model (SM) admit the existence
of new heavy gauge bosons [1] in both the neutral ($Z^{\prime}$) and  the charged
 ($W^{\pm\prime}$) sectors. From time to time the question of why the top quark is so heavy
is addressed to the physics community. Some of the modern theoretical models can offer
clarification of this problem within the new additional gauge interactions related to new gauge
bosons and fermions of the third and even the fourth generations.
The models operating in this direction carry out the extension of the gauge group $SU(N)$ up to
$SU(N)\times SU(N)$ gauge structure [2,3]. Due to spontaneous breaking of
$SU(N)\times SU(N)$ to its diagonal subgroup the relevant generators correspond to the set
of massive $SU(N)$ gauge bosons interacting with fermions of any generations with
different coupling strengths. The models of the $SU(N)\times SU(N)$ type predict, in fact,
the existence of the $Z^{\prime}$-boson interacting mainly and efficiently with fermions
of the third generation. The simplest theories are based on the extensions of type
$SU(2)_{l}\times SU(2)_{h}$ for weak interactions: the first two generations of fermions
are correlated with the weak gauge group $SU(2)_{l}$, while the third generation "feels" another,
"stronger" gauge structure $SU(2)_{h}$.

Notice that in some models containing the extended gauge group $SU(2)_{l}\times SU(2)_{h}$
the lower bound on the $Z^{\prime}$-boson mass is given by the scale of
1.0-1.5 GeV [4]. In this paper, we restrict our consideration to the relatively light $SU(2)$
$Z^{\prime}$-bosons of the order $O$(1 GeV). From the phenomenological point of view
the interest in these bosons is caused by the possibility of their identification at
the Fermilab Tevatron with the energy $\sqrt s\simeq$ 2 TeV and the $pp$- large hadron
collider at the CERN with $\sqrt s\simeq$ 14 TeV.

The models in which the precision data on electroweak interactions admit the existence
of only light new gauge bosons $Z^{\prime}$ and $W^{\pm\prime}$ are concerned with
the extended models with the technicolour [3].  The electroweak symmetry that breaks
down the
technicolour structure is characterized at least by two stages needed to transition from
nonbreaking symmetry at high energies to the low-energy electromagnetic gauge structure.
The idea is that on some scale $u$  the symmetry of two gauge groups $SU(2)$
($SU(2)_{l}$ and $SU(2)_{h}$) is breaking to their diagonal subgroup. The break
down of the rest electroweak symmetry originates at the scale $v < u$ ($v$= 246 GeV).

The $Z-Z^{\prime}$ mixing effects are defined by the angle $\xi$ within the well-known
relation between the eigenstates $Z_{1}$ and $Z_{2}$ with the masses $m_{Z_{1}}$ and
$M_{Z_{2}}$, respectively [1]
\begin{eqnarray}
\label{e1}
{Z_{1}\choose Z_{2}}={\cos\xi\,\, \sin\xi\choose -\sin\xi\,\,\cos\xi}{Z\choose Z^{\prime}}\, ,
\end{eqnarray}
and
\begin{eqnarray}
\label{e2}
\vert\xi\vert={\arctan}\left (\frac{m^{2}_{Z}-m^{2}_{Z_{1}}}{M^{2}_{Z_{2}}-m^{2}_{Z}}\right )
^{\frac{1}{2}},
\end{eqnarray}
where $m_{Z}$ is the $Z$-boson mass. Very often the effects due to the $\xi$-angle are neglected
because of its rather small value. Nevertheless, the estimation of the absolute value of $\xi$
is an important task by itself.

One of the main questions is the signature with which new gauge bosons should be displayed in the
experiments. The promising modes of new heavy
gauge boson decays as well as the proposals for the experimental restriction on the masses
of $Z^{\prime}$- and  $W^{\pm\prime}$-bosons can be found in [5-10].
 The Tevatron Run I data can allow one to estimate the lower bound on the $Z^{\prime}$-boson mass at
the level of $M_{Z^{\prime}}>$  650 GeV as the narrow resonance with the width
$\Gamma_{Z^{\prime}}= 0.12~M_{Z^{\prime}}$ [11]. In fact, the partial decay widths in
[5-10] are defined with an accuracy up to new coupling constants which (as the
rule) are unknown, e.g., the vector and axial-vector couplings of $Z^{\prime}$ with the quark fields.
The most promising channels to observe $Z^{\prime}$-bosons are their decays into quarks and
antiquarks of the third generation ($b\bar b, \,t\bar t$) [5-8] or even the pairs of leptons
and antileptons in the current experiments at the Tevatron or forthcoming experiments
at the LHC (e.g., with the help of the ATLAS detector[12]). Thus, the hadron colliders
$Z^{\prime}$-boson discovery potential depends on the couplings of $Z^{\prime}$ with quarks and leptons.

The partial widths in the decays mentioned above are on the scale level of O(1 GeV).\\
We investigate the radiative decays of an  extra gauge boson $Z^{\prime}$ with
production of the standard $Z$-boson ($Z^{\prime}\rightarrow\gamma Z$), because the
monochromatic photon could be efficiently displayed in the experiments, while the
accompanying processes like $Z^{\prime}\rightarrow \gamma +$ pseudoscalar mesons will be suppressed.
The decay $Z^{\prime}\rightarrow \gamma  \gamma$ is forbidden because of the Bose-symmetry.
For now we assume that there will be possible to observe a peak in the photon spectrum of the
inclusive process  $Z^{\prime}\rightarrow \gamma + all$. Another important feature of the
decay  $Z^{\prime}\rightarrow \gamma Z$ is the possibility to get the restriction on the
couplings between $Z^{\prime}$ and quarks. The relative width of the decay
$Z^{\prime}\rightarrow \gamma Z$ compared to
$Z^{\prime}\rightarrow \mu^{+}\mu^{-}, \, Z^{\prime}\rightarrow \tau^{+}\tau^{-}$
decays is expected on the level of about $10^{-5}-10^{-4}$, which corresponds to
the partial decay width $\Gamma (Z^{\prime}\rightarrow \gamma Z)\sim O (1~GeV)$.
The decay $Z^{\prime}\rightarrow \gamma Z$ has the peculiarity of the
enhancement effect because of the factor
$G_{F}\,M^{4}_{Z^{\prime}}/m^{2}_{Z}$ due to longitudinal polarization of the
$Z$-boson ($G_{F}$ is the Fermi weak constant). Here, for simplicity, one can
neglect the effects of mixing
between the physical states $Z^{\prime}$ and $Z$, and one can take into account only
axial couplings of $Z^{\prime}$ with the quarks in the loop. We believe that all the quarks
may have the identical $U^{\prime}(1)$ charge. The mass of the $Z^{\prime}$-boson is
unknown and it is generated by the own scalar singlet in the framework of the
$SU(2)\times U(1)$ group. \\

\centerline  {\bf 2. General properties of $SU(2)$  $Z^{\prime}$-bosons }

Consider the physical model containing weak interactions governed by the pair of
$SU(2)$ gauge groups: $SU(2)_{h}\times SU(2)_{l}$ [9]. The $SU(2)_{h}$ group is
responsible for weak interactions with heavy leptons and quarks, and  the
left fermions transform as doublets while the right fermions as  singlets.
The group $SU(2)_{l}$ is related to the first two generations of leptons and quarks
with the charges identical to those which come from the standard representation.
The extended gauge group $SU(2)_{h}\times SU(2)_{l}$  breaks down to its diagonal
subgroup $SU(2)_{L}$ on the scale $u$ (which is yet unknown) by some scalar field
$\sigma$ with the charge $SU(2)_{h}\times SU(2)_{l}\times U(1)_{Y}$
$$\sigma\sim (2\, ,2)_{0}\,\,\, , \langle\sigma\rangle ={u\,\,\,\,\, 0\choose
0\,\,\,\,\, u} \, .$$
The symmetry breaking $SU(2)_{L}\times U(1)_{Y}\rightarrow U(1)_{em}$ takes place
because of nonzero value of the vacuum expectation value of the scalar field
$v= 246$ GeV.\\
The set of the gauge fields looks like
\begin{eqnarray}
\label{e3}
A^{\mu}=\sin\Theta_{W}\,(\cos\phi\,W_{3h}^{\mu}+\sin\phi\,W_{3l}^{\mu})+\cos\Theta_{W}\,X^{\mu}\, ;
\end{eqnarray}
\begin{eqnarray}
\label{e4}
Z^{\mu}_{1}=\cos\Theta_{W}\,(\cos\phi\,W_{3h}^{\mu}+\sin\phi\,W_{3l}^{\mu})-\sin\Theta_{W}\,X^{\mu}\, ;
\end{eqnarray}
\begin{eqnarray}
\label{e5}
Z^{\mu}_{2}=-\sin\phi\,W_{3h}^{\mu}+\cos\phi\,W_{3l}^{\mu}
\end{eqnarray}
with the $U(1)_{em}$ group generator $Q=T_{3h}+T_{3l}+Y$ as an operator of the electric charge.
Notice that the covariant derivative is getting longer with the additional term related to
$Z^{\prime}$-boson
\begin{eqnarray}
\label{e6}
D^{\mu}=\partial^{\mu}-i\,\frac{g}{\cos\Theta_{W}}\,Z^{\mu}_{1}\,(T_{3}-Q\,\sin^{2}\Theta_{W})-
i\,Z^{\mu}_{2}(-g_{h}\,\sin\phi\,T_{3h}+g_{l}\,\cos\phi\,T_{3l})\, ,
\end{eqnarray}
where $T_{3}=T_{3h}+T_{3l}$, $\phi$ is the additional mixing angle, and the gauge coupling constants
$g_{h}$ and $g_{l}$ have the following form:
$$ g_{h} = \frac{g}{\cos\phi}\,\, ,\,\,\, g_{l} = \frac{g}{\sin\phi} . $$

The effects of the mixing $Z_{1}^{\mu} - Z_{2}^{\mu}$ are defined by the angle $\phi$, and
in the leading order over $1/x = v^{2}/u^{2}$ there is the following field superposition [9]:
\begin{eqnarray}
\label{e7}
{Z^{0}\choose Z^{\prime}}={1\,\,\,\,\, -\frac{\cos^{3}\phi\,\sin\phi}{x\,\cos\theta}\choose
\frac{\cos^{3}\phi\,\sin\phi}{x\,\cos\theta}\,\,\,\,\,1}{Z_{1}\choose Z_{2}}\, .
\end{eqnarray}\\

\centerline  {\bf 3. The model}

Our approach could be applied almost to all the models relevant to the
$Z^{\prime}$-boson (e.g., $E_{6}$ superstring model, etc.); however, we
use the extended $SU(2)_{h}\times SU(2)_{l}$ model with the fermion loops
containing $up$- and $down$ - quarks including the quarks of the fourth generation.
For now we assume that the couplings of $Z^{\prime}$ and $Z$ with quarks $q$ are
defined through the following Lagrangian density:
$$ -L = g_{Z}\,\sum_{q}\,\bar q (v_{q} - a_{q}\,\gamma_{5})\,
\gamma^{\mu}\,q\,Z_{\mu} + g^{\prime}_{Z}\,\sum_{q}\,\bar q
(v^{\prime}_{q} - a^{\prime}_{q}\,\gamma_{5})\,
\gamma^{\mu}\,q\,Z^{\prime}_{\mu},$$
$v_{q}\, ,a_{q}$ and $v^{\prime}_{q}\, , a^{\prime}_{q}$ - are the set
of vector and
axial-vector coupling constants for $Z$ and $Z^{\prime}$, respectively.
The mixing $Z^{\prime}$-$Z$ effect is given by the term $\sim \xi(M^{2}_{Z^{\prime}}-
m^{2}_{Z})Z^{\prime}_{\mu}Z^{\mu}$ in the total Lagrangian density.
Without loss of generality, the angle of the mixing
$Z^{\prime}$-$Z$ is not included because of its rather small value.
Suppose the decay $Z^{\prime}\rightarrow \gamma Z$ in  the lowest order on the
coupling constant is given by the loop diagrams containing the quarks  mainly
of the third and the fourth generations. The matrix element of this process is
\begin{eqnarray}
\label{e8}
M=\frac{1}{16\,\pi^2}\,e\,g_{Z}\,g_{Z^{\prime}}\,F\,\epsilon_
{\mu\nu\alpha\beta}\,\epsilon^{\mu}_{\gamma}\,\epsilon^{\nu}_{Z^{\prime}}\,
\epsilon^{\alpha}_{Z}\,p^{\beta}\, ,
\end{eqnarray}
where $e=\sqrt {4\,\pi\,\alpha}$ ($\alpha$ is the effective electromagnetic coupling constant
to be evaluated on the scale $M_{Z^{\prime}}$, $\alpha\sim$ 1/128),
$g_{Z}=g/cos\Theta_{W}$, $F=\sum_{q:b,t,...}e_{q}\,T_{3q}$ ($e_{q}$ is the electric quark charge,
$T_{3q}$ is the third component of the weak isospin), $g_{Z^{\prime}}$ being the free parameter,
while $\epsilon^{\mu}_{\gamma}\, , \epsilon^{\nu}_{Z^{\prime}}\, ,\epsilon^{\alpha}_{Z}$ are the
wave functions of $\gamma$-quantum, $Z^{\prime}$-, $Z$- bosons, respectively; $p^{\mu}$ is the
4-momentum of the $\gamma$-quantum.
In the Grand Unification models $g_{Z^{\prime}}$ is related to $g_{Z}$ in the following way
(see, e.g., [13])
\begin{eqnarray}
\label{e9}
g_{Z^{\prime}}=\sqrt{\frac{5}{3}\,\lambda}\,\sin\Theta_{W}\,g_{Z}\simeq 0.62\,g_{Z}\, ,
\end{eqnarray}
where $\lambda\sim O(1)$.
Making the replacement with $\epsilon_{Z}^{\mu}\rightarrow  q^{\mu}/m_{Z}$ in (\ref{e8})
one can
take into account the longitudinal polarization of $Z$-boson in the leading order in
$1/m_{Z}$ ($q^{\mu}$ is the 4-momentum of $Z$).\\
The absolute width of the decay $Z^{\prime}\rightarrow \gamma Z$ is given by
\begin{eqnarray}
\label{e10}
\Gamma(Z^{\prime}\rightarrow \gamma Z)=\frac{5\,\alpha\lambda\,\sin^{2}\Theta_{W}\,G_{F}^{2}}
{96\,\pi^4}\,F^2\,m^{2}_{Z}\,M^{3}_{Z^{\prime}}\left (1-\frac{m^{2}_{Z}}{M^{2}_{Z^{\prime}}}\right )
^{3}\, .
\end{eqnarray}
There is a peculiarity in (\ref{e10}) indicating the independence of the decay width
relevant to quark masses due to the suppression factor  $\sim (m_{q}/M_{Z^{\prime}})^{2}$ taking into
account the experimental restriction on the lower bound of the $Z^{\prime}$-boson mass.

%%%%%%%%%%%%%%%%%%%%%%%%%%%%%%%%%%%%%%%%%%%%%%%%%%%%%%%%%
\begin{center}
\begin{figure}[h]
\resizebox{12cm}{!}{\includegraphics{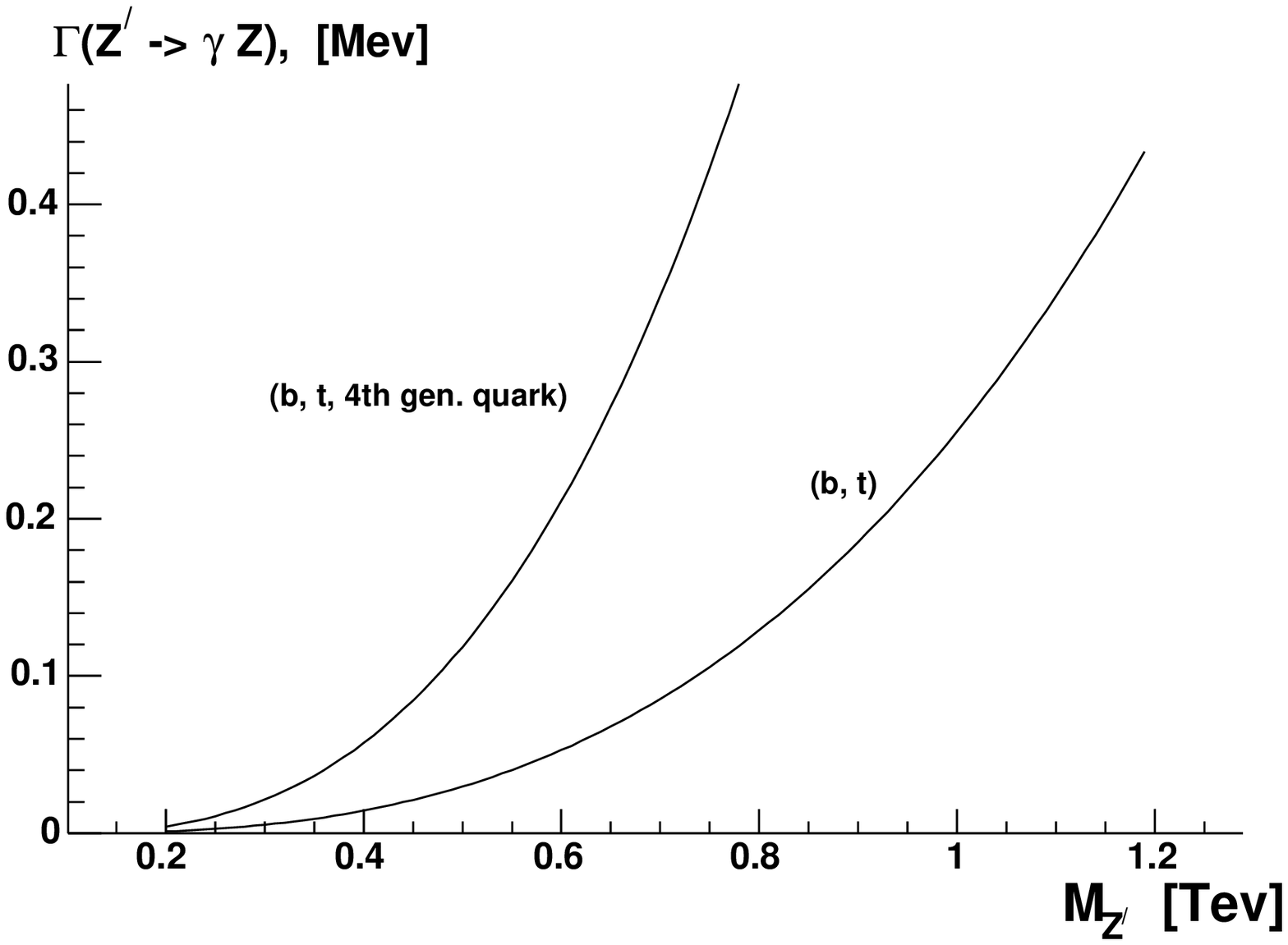}}

Fig.1 Decay width $\Gamma(Z^{\prime}\rightarrow \gamma Z)$
as a function of $M_{Z^{\prime}}$ with the contribution due to only
$b$- and $t$-quarks in the loop  and taking into account the additional
contribution extended by the quarks of the fourth generation.

\end{figure}
\end{center}

%%%%%%%%%%%%%%%%%%%%%%%%%%%%%%%%%%%%%%%%%%%%%%%%%%%%%%%%%%

In Fig.1 we present the results of calculation of
$\Gamma(Z^{\prime}\rightarrow \gamma Z)$ as
a function of $M_{Z^{\prime}}$ with the contribution due to only $b$- and
$t$-quarks in the loop and taking into account the additional
contribution extended by the quarks of the fourth generation. It is worthy to note
that the contribution, thanks to quarks of the fourth generation, leads to the
4 times enhancement of $\Gamma(Z^{\prime}\rightarrow \gamma Z)$.
The parameter $\lambda$ is equal to 1 in the calculations.\\
To use the results of the data analysis carried out at the LEP [14,15] and the
Tevatron [9], based on the study of the influence of the low-energy effects of
quark-lepton contact interactions toward the lepton pair production, we make
use of the processes $Z^{\prime}\rightarrow\tau^{+}\tau^{-}$ and
$Z^{\prime}\rightarrow\mu^{+}\mu^{-}$ as the normalized ones to the decay
$Z^{\prime}\rightarrow \gamma Z$ we are interested in.

The $l \bar l$ leptonic pair decay width of the $Z^{\prime}$-boson is given by the formula
\begin{eqnarray}
\label{e11}
\Gamma(Z^{\prime}\rightarrow \bar l l)=\frac{g_{Z^{\prime}}^{2}\,M_{Z^{\prime}}}{12\,\pi}\,
\{v^{\prime\,2}_{l}\,(1+2\,y)^{2}+a^{\prime\,2}_{l}\,(1-4\,y)\}\sqrt{1-4\,y}\, ,
\end{eqnarray}
where $y=m_{l}^{2}/M_{Z^{\prime}}^{2}$, $m_{l}$ means the mass of the lepton, $v^{\prime}_{l}$
and  $a^{\prime}_{l}$ are vector and axial-vector interaction constants of  $Z^{\prime}$
with leptons. The parameters $v^{\prime}_{l}$  and  $a^{\prime}_{l}$ reflect the chiral
properties of the $Z^{\prime}$-boson interaction to leptons as well as the relative
strength of these interactions
$$- L_{Z^{\prime}l} = g_{Z^{\prime}}\,\sum_{l}\,\bar l(v^{\prime}_{l}- a^{\prime}_{l}\gamma_{5})
\,\gamma^{\mu}\,l\,Z^{\prime}_{\mu}\, .$$
The ratio between the widths of the decays $Z^{\prime}\rightarrow\gamma Z$ and
$Z^{\prime}\rightarrow\bar l l$ can be defined as
\begin{eqnarray}
\label{e12}
R(Z^{\prime}\rightarrow \gamma Z/\bar l l) \equiv\frac{BR(Z^{\prime}\rightarrow \gamma Z)}
{BR(Z^{\prime}\rightarrow\bar l l)}
=\frac{3\,\sqrt{2}\alpha\,G_{F}\,F^{2}\,M^{2}_{Z^{\prime}}}
{64\,\pi^3\,(v^{\prime\,2}_{l}+a^{\prime\,2}_{l})}\,{\left (1-\frac{m^{2}_{Z}}
{M^{2}_{Z^{\prime}}}\right )}^{3} .
\end{eqnarray}
Notice that the sum $v^{\prime\,2}_{l}+a^{\prime\,2}_{l}$ in (\ref{e12}) may be different
depending of the type of leptons. In [8],  the following
normalization relation $v^{\prime\,2}_{\tau}+a^{\prime\,2}_{\tau}=0.5$ in the case of
$\tau-$leptons was used. This allows us to apply this
condition to the normalization process $Z^{\prime}\rightarrow\tau^{+}\tau^{-}$.
In Fig.2,  we plot the ratio $R(Z^{\prime}\rightarrow \gamma Z/\bar l l)$  as a function
of the mass  $M_{Z^{\prime}}$ in the $\tau$-lepton channel ($l = \tau$) taking into account
the contribution from only $b$- and $t$-quarks and with the additional loop containing the
quarks of the fourth generation.  The resulting calculation of
$R(Z^{\prime}\rightarrow \gamma Z/ \mu^{+}\,\mu^{-})$ (in $\mu$-meson channel) does not
differ essentially in comparison with $R(Z^{\prime}\rightarrow \gamma Z/\tau^{+}\tau^{-})$
if one believes in the following coupling constant relation:
$v^{{\prime}^{2}}_{\mu} + a^{{\prime}^{2}}_{\mu} = v^{{\prime}^{2}}_{\tau} +
a^{{\prime}^{2}}_{\tau}$.

%%%%%%%%%%%%%%%%%%%%%%%%%%%%%%%%%%%%%%%%%%%%%%%%%%%%%%%%%
\begin{center}
\begin{figure}[h]
\resizebox{12cm}{!}{\includegraphics{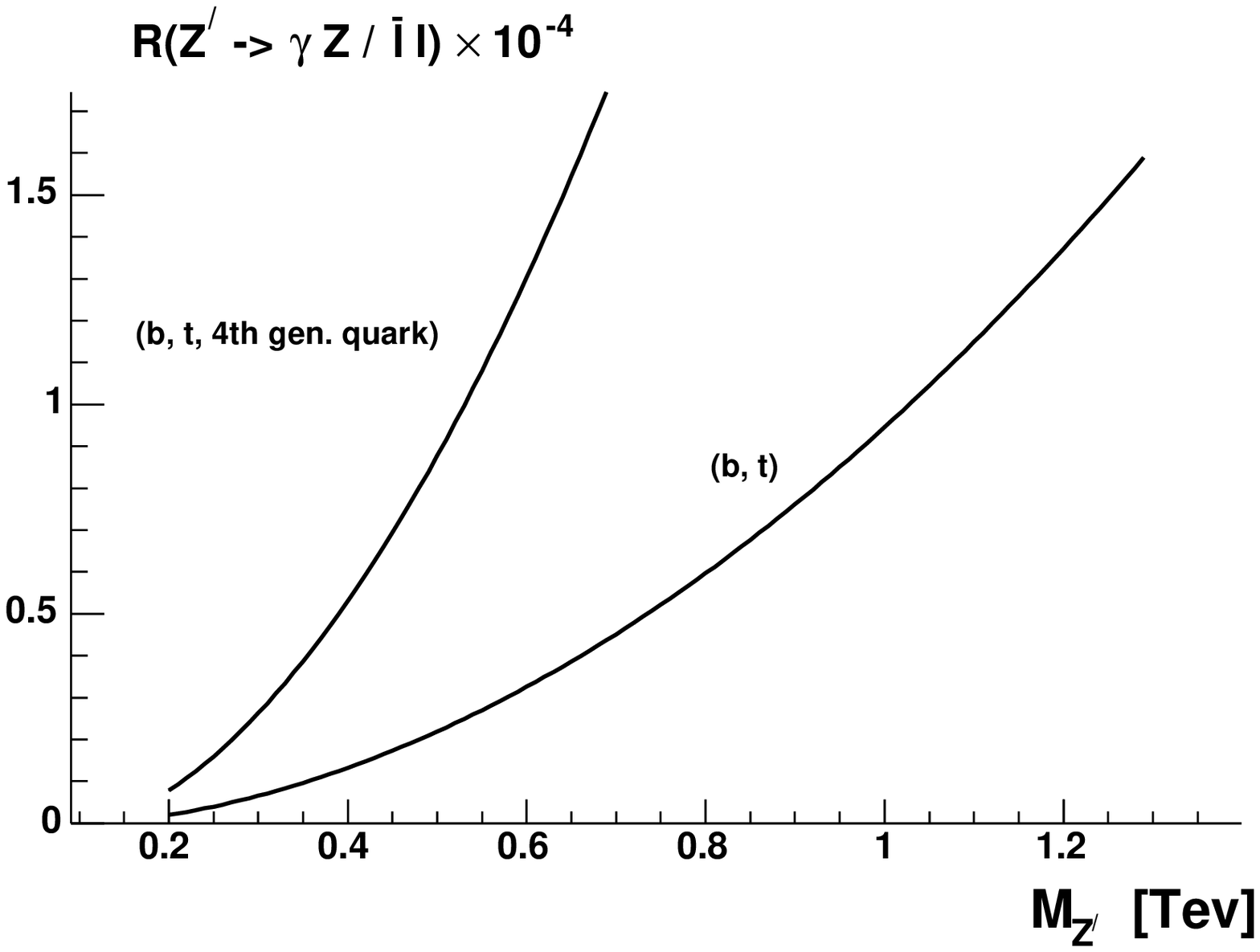}}

Fig.2 The ratio $R(Z^{\prime}\rightarrow \gamma Z/\bar l l)$ as a function
of the mass  $M_{Z^{\prime}}$ in the $\tau$-lepton channel ($l = \tau$) taking
into account the contribution from only $b$- and $t$-quarks and with the additional
loop containing the quarks of the fourth generation.

\end{figure}
\end{center}

%%%%%%%%%%%%%%%%%%%%%%%%%%%%%%%%%%%%%%%%%%%%%%%%%%%%%%%%%%

To estimate the lower bound on $R(Z^{\prime}\rightarrow \gamma Z/\tau^{+}\tau^{-})$,
we  use the results [9] of the contact 4-fermion interaction analysis in
$e^{+}e^{-}\rightarrow \tau^{+}\tau^{-}$. The mass $M_{Z^{\prime}}$ is given by
the mass scale parameter $\Lambda$ which is typical of the scale of new physics (NP)
\begin{eqnarray}
\label{e13}
M_{Z^{\prime}}=\Lambda\,\frac{\sqrt\alpha}{2\,\sin\Theta_{W}}\, ,
\end{eqnarray}
where $\Lambda >$ 3.8 TeV [14] or $\Lambda >$ 3.9 TeV [15]. Our numerical result is
$R(Z^{\prime}\rightarrow \gamma Z/\tau^{+}\tau^{-})> 1.3\times 10^{-5}$. Using
the resulting analysis of the CDF Collaboration [9]  and taking into
consideration the normalization process
$Z^{\prime}\rightarrow \mu^{+}\mu^{-}$, we get the following estimation
$R(Z^{\prime}\rightarrow \gamma Z/\mu^{+}\mu^{-})> 1.4\times 10^{-5}\cot^{2}\phi$, if
$v^{\prime\,2}_{\mu}+a^{\prime\,2}_{\mu}=0.5$ (see the $\tau$-lepton case as well).\\
To study  the $Z^{\prime}$-boson decays with the production of the
third generation quarks (Q), $Z^{\prime}\rightarrow \bar b b$,
$Z^{\prime}\rightarrow \bar t t$ (normalized decays), we can find lower
values for $R(Z^{\prime}\rightarrow \gamma Z/\bar Q Q)$
\begin{eqnarray}
\label{e14}
R(Z^{\prime}\rightarrow \gamma Z/\bar b b)\equiv\frac{BR(Z^{\prime}\rightarrow \gamma Z)}
{BR(Z^{\prime}\rightarrow \bar b b)} = (0.13 - 1.60)\times 10^{-6}
\end{eqnarray}
in the mass region 0.2 TeV$ < M_{Z^{\prime}} <$ 1.0 TeV, and
\begin{eqnarray}
\label{e15}
R(Z^{\prime}\rightarrow \gamma Z/\bar t t)\equiv\frac{BR(Z^{\prime}\rightarrow \gamma Z)}
{BR(Z^{\prime}\rightarrow \bar t t)} = (0.61 - 1.80)\times 10^{-6}
\end{eqnarray}
for 0.4 TeV$ < M_{Z^{\prime}} <$ 1.0 TeV, which is natural because the $\bar Q Q$-channel,
especially $\bar b b$, is more probable to be compared with leptonic decays.

Relation (\ref{e2}) allows us to estimate $M_{Z^{\prime}}$ as a function of
the mixing angle $\xi$ and the mass parameter  $\Delta_{M} = M_{Z^{\prime}} - M_{Z_{2}}$
\begin{eqnarray}
\label{e16}
M_{Z^{\prime}}\simeq\Delta_{M} + \frac{1}{\vert\xi\vert}\,\sqrt{m^{2}_{Z} - m^{2}_{Z_{1}}}\,\,
,\,\,\,\vert\xi\vert\neq 0\, .
\end{eqnarray}
Following [16] we keep in mind the mixing between $Z^{\prime}$ and $Z$ in the form of
$\rho_{mix}$ - factor in the function
\begin{eqnarray}
\label{e17}
\rho = \rho_{top}\cdot\rho_{mix} \, ,
\end{eqnarray}
occurring in the interaction constants of the SM due to the $Z^{\prime} -Z$ mixing.
In formula (\ref{e17}) one has
\begin{eqnarray}
\label{e18}
 \rho_{top} = \frac{1}{1-\delta\rho_{top}}\,\,\,  ,\,\,\, \delta\rho_{top}\simeq \frac{
 3\,G_{F}}{8\,\sqrt{2}\,\,\pi^{2}}\,\cdot m^{2}_{top}\simeq 0.01
\end{eqnarray}
which reflects the one-loop correction due to the top-quark contribution, while
\begin{eqnarray}
\label{e19}
 \rho_{mix} = 1 + \sin^{2}\xi\,\left (\frac{ M^{2}_{Z_{2}}}{m^{2}_{Z_{1}}} - 1\right )\,
\end{eqnarray}
with  $\rho_{0} = (m_{W}/m_{Z}\,\cos\Theta)^{2} =1$ in the SM. Then, relation
(\ref{e16}) is
\begin{eqnarray}
\label{e20}
M_{Z^{\prime}}\simeq\Delta_{M} + \frac{m_{Z}}{\vert\xi\vert}\,\sqrt{1 -
\frac{1}{\rho_{mix}}}\, .
\end{eqnarray}
In this way, we can obtain the absolute value of $\vert\xi\vert$
\begin{eqnarray}
\label{e21}
\vert\xi\vert = \frac{m_{Z}}{C^{exp} - \Delta_{M}}\,\sqrt{1 - \frac{1}{\rho_{mix}}}\, ,
\end{eqnarray}
where $C^{exp}$ is the minimal value of the massive scale parameter of NP, which
has been estimated at the LEP [14,15] and the Tevatron [11] under the contact four-fermion
interaction analysis in $e^{+}e^{-}\rightarrow Z^{\prime}\rightarrow \tau^{+}\tau^{-}$ and
$\bar p p\rightarrow Z^{\prime}\rightarrow e^{+}e^{-}\, ,\mu^{+}\mu^{-}$,
respectively: $C^{exp}_{LEP} = 0.355$ TeV [14], $C^{exp}_{LEP} = 0.365$ TeV [15],
$C^{exp}_{CDF} = 0.345\cdot\cot\phi $ TeV (electrons channel) [11] and
$C^{exp}_{CDF} = 0.380\cdot\cot\phi $ TeV (muons channel) [11].\\
To conclude this section, we derive an estimation on the mixing angle $\xi$.
Actually, one can use the fact that $\rho_{mix}$ could
be extracted from (\ref{e17}), where the $\rho_{top}$-factor is already known, and
the full factor $\rho$ enters in its turn in the redefined constant $g_{Z}$ of the
interaction of the
$Z$-boson with the fermions, $g_{Z} = (g/\cos\Theta_{W})\cdot \rho^{1/2}$ or even in
the updated value of $\sin^{2}\Theta_{W}$
$$\sin^{2}\Theta_{W} = \frac{1}{2} - \left [\frac{1}{4} - \frac{\pi\,\alpha(m_{Z})}
{\sqrt{2}\,G_{F}\,\rho\,m_{Z}^{2}}\right ]^{\frac{1}{2}}\, .$$
The numerical analysis shows that the upper limit on $\vert\xi\vert$, obtained within the
CDF data [11] is comparable with those based on the LEP data [14,15] only under the
following condition $\sin\phi\simeq\cos\phi$. In this case our estimations lead to the following
restriction on $\vert\xi\vert$:
\begin{eqnarray}
\label{e22}
0 < \vert\xi\vert < 5\times 10^{-2}
\end{eqnarray}
at $ 0 < \delta < 0.05$, where $\delta = \rho_{mix} -1$. The increase in the mixing angle
$\phi$ up to the physically grounded one $\sin\phi = 0.85 $ [9] leads to the insignificant
rise of the upper limit of $\vert\xi\vert$, $0 < \vert\xi\vert < 6\cdot 10^{-2}$, in the
above-mentioned interval for $\delta$.

The ratio between two scales $u/v$ of break down of the initial symmetry in the
$SU(2)_{h}\times SU(2)_{l}$ - model has the following lower bound:
\begin{eqnarray}
\label{e23}
\frac{u}{v}> \Lambda \,\frac{\cos^{2}\phi\,\sqrt{\alpha}}{2\,\sin\Theta_{W}\,m_{W}} .
\end{eqnarray}
To get the estimation on the upper limit of the scale $u$ one can take into consideration the
fact that in the framework of the model $SU(2)_{h}\times SU(2)_{l}$ the minimal value of
the full width $\Gamma_{Z^{\prime}}$ of the $Z^{\prime}$-boson decay is achieved at
$\sin\phi\simeq$ 0.8 [9]. Moreover, $\sin\phi$ is restricted by the value
$\sim$ 0.85, where the sharp rise of $\Gamma_{Z^{\prime}}$ begins because of the
dominant contribution of the gauge coupling constant $g_{h}=g/(1- \sin^{2}\phi)^{1/2}$
($g=e/\sin\Theta_{W}$). Thus, the values $\sin\phi$ can be restricted in the window
 $0.6\leq\sin\phi\leq 0.85$, where the contribution of the constant
$g_{l}=g/\sin\phi$ ($1/g^{2}=1/g^{2}_{h}+ 1/g^{2}_{l}$) becomes insignificant compared
to $g_{h}$. The numerical estimations indicate $1.16\leq (u/v)\leq 2.98$, where the lower
bound corresponds to $\sin\phi$=0.85 while the upper limit is achieved at $\sin\phi$=0.6.\\

\centerline {\bf 4. Results}

In  conclusion let us formulate the main results.
In the paper, the radiative decays of new extra gauge bosons the $Z^{\prime}$
accompanying the production of the $Z$-boson are investigated in the framework
of the extended $SU(2)_{h}\times SU(2)_{l}$ gauge model.
We investigated the induced amplitude of the radiative decay
$Z^{\prime}\rightarrow Z\gamma$ with the finite value of the relative decay width
which can be measurable and thus, the relations between
$Z^{\prime}$ and fermions should be clarified. The results presented here are based on the
consideration of the sensitivity of  $M_{Z^{\prime}}$ to $\Lambda$ in the framework
of the contact quark-lepton interaction. The resulting truth depends mainly
on the mixing angle $\phi$, and the interaction constants $g_{Z^{\prime}}$, $v^{\prime}_{l}$
and $a^{\prime}_{l}$. The results obtained give a promising chance for an indirect
observation of the $Z^{\prime}$-bosons in their decays in lepton-antilepton pairs
or even in their resonances in the photonic spectra because of the
radiative decays. The extended model  $SU(2)_{h}\times SU(2)_{l}$ admits the strengthening
interaction between the third generation fermions with the extended gauge sector.
In this case, the study of the final states $\mu^{+}\mu^{-}\, ,\bar b b\, ,\bar t t$ is
most important. However, the decay modes $Z^{\prime}\rightarrow \bar t t$  and
$Z^{\prime}\rightarrow \bar b b$ are rather difficult for observation because of their
big invariant mass, e.g., for $ \bar t t$-pair, as well as the large QCD background in
the case of $ \bar b b$-pair. The most probable channel is the production of two
$\tau$-leptons, $\bar p p(pp)\rightarrow Z^{\prime}\rightarrow \tau^{+}\tau^{-}+X$.
The precise measurement of the angle distributions for the pairs of leptons and antileptons
($\mu^{+}\mu^{-}$ or $\tau^{+}\tau^{-}$) will allow one to clarify the very important and
instructive information on the structure
of the interaction constants between $Z^{\prime}$ and fermions, and hence, to get
an indirect confirmation of
existence of a new heavy gauge boson, or even a set of gauge bosons, as well as to restrict
the domain of their investigation, e.g., over their  masses.\\
The resulting CDF data analysis for the "gauge" process
$Z^{\prime}\rightarrow\mu^{+}\mu^{-}$ can allow one to define the lower bound
on the decay probability $R(Z^{\prime}\rightarrow\gamma Z/\mu^{+}\mu^{-})>
1.4\times 10^{-5}\cot^{2}\phi$ in the case
$v^{{\prime}^{2}}_{\mu} + a^{{\prime}^{2}}_{\mu} = v^{{\prime}^{2}}_{\tau} +
a^{{\prime}^{2}}_{\tau}\simeq 0.5$ we have proposed,  and to give the estimation on the mixing angle
$\phi $ of the extended gauge group $SU(2)_{h}\times SU(2)_{l}$.
In addition, the quarks of the fourth generation can lead to 4 times
increase in the decay width $\Gamma(Z^{\prime}\rightarrow\gamma Z)$.
The CDF data in application to four-fermion contact interaction at
$\cot\phi\simeq 1$ in order to estimate the mixing angle $\xi$ for the system
$Z^{\prime}-Z$ are comparable with the data given by LEP, thus leading to the following
restriction $\vert\xi\vert < 5\times 10^{-2}$. Here the special role of the group
$SU(2)_{h}$ relevant to top-quarks (and hence $\tau$-leptons) shows up, which
is significant to current experiments at the Tevatron. We emphasize that
the group $SU(2)_{h}$ is just a conductor of weak interactions where
top-quarks and $\tau$-leptons are involved. As noted in Chapter 3, the LEP data
relevant to decays
 $Z^{\prime}\rightarrow\tau^{+}\tau^{-}$ and
$Z^{\prime}\rightarrow \bar b b$, $Z^{\prime}\rightarrow \bar t t$ are taken into
account when we carried out the comparison  between the Tevatron and the LEP data analyses.

It turns out that a  probable investigation of  radiative decays  of $Z^{\prime}$-bosons at
the acting hadron collider - the Fermilab Tevatron, would allow one to clarify, e.g.,
the question of
the efficiency of the gauge group $SU(2)_{h}\times SU(2)_{l}$, and to give the estimation
on the vector and axial-vector coupling constants of $Z^{\prime}$-boson interaction with
fermions.

Undoubtedly, for a further deeper study of $Z^{\prime}$-bosons the most urgent
step would be the investigation of  decays like $Z^{\prime}\rightarrow \tau^{+}\tau^{-}+X$
with the restriction on the parameter $\Lambda$ itself, and  hence, on the mass $M_{Z^{\prime}}$.
Each of $ \tau$-leptons (in the final state) decays into the hadrons
($BR(\tau\rightarrow $hadrons)$\simeq$ 0.65) or even into charge leptons and
neutrinos $\tau^{+}\tau^{-}\rightarrow e^{+} e^{-}\nu\, ,\,\,
\tau^{+}\tau^{-}\rightarrow\mu^{+}\mu^{-}\nu\, ,\,\,
\tau^{+}\tau^{-}\rightarrow e^{+}\mu^{-}\nu (e^{-}\mu^{+}\nu )$
($BR(\tau\rightarrow $ leptons)$\simeq$ 0.35). Therefore, new data on the processes like
 $\bar p p \rightarrow Z^{\prime}\rightarrow \tau^{+}\tau^{-}+X$ at the Tevatron with the
energy $\sqrt{s}\simeq $ 2 TeV would be very important.\\

\centerline {\bf Acknowledgements}
I am grateful to G. Bellettini, A.N. Sissakian, and Yu.A. Boudagov for useful discussions,
and to G. Koriauli for help in numerical calculations.

\end{document}